\documentclass[11pt,a4paper]{article}
\usepackage{jheppub}

\def\bit{\begin{itemize}}
\def\eit{\end{itemize}}
\def\ben{\begin{enumerate}}
\def\een{\end{enumerate}}
\def\bed{\begin{description}}
\def\eed{\end{description}}

\def\k{\kappa}
\def\l{\lambda}

\def\q{\quad}

\def\mff{\mu_{\rm eff}}

\def\cmg{\, {\rm cm^2/g} }
\def\half{\frac{1}{2}\,}
\def\third{\frac{1}{3}\,}

\def\lsim{\raise0.3ex\hbox{$<$\kern-0.75em\raise-1.1ex\hbox{$\sim$}}}
\def\gsim{\raise0.3ex\hbox{$>$\kern-0.75em\raise-1.1ex\hbox{$\sim$}}}

\let\jnfont=\rm
\def\NPB#1,{{\jnfont Nucl.\ Phys.\ B }{\bf #1},}
\def\PLB#1,{{\jnfont Phys.\ Lett.\ B }{\bf #1},}
\def\EPJC#1,{{\jnfont Eur.\ Phys.\ Jour.\ C }{\bf #1},}
\def\PRD#1,{{\jnfont Phys.\ Rev.\ D }{\bf #1},}
\def\PRL#1,{{\jnfont Phys.\ Rev.\ Lett.\ }{\bf #1},}
\def\MPLA#1,{{\jnfont Mod.\ Phys.\ Lett.\ A }{\bf #1},}
\def\JPG#1,{{\jnfont J.\ Phys.\ G}{\bf #1},}
\def\CTP#1,{{\jnfont Commun.\ Theor.\ Phys.\ }{\bf #1},}
\def\JHEP#1,{{\jnfont JHEP \ }{\bf #1},}
\def\NPPS#1,{{\jnfont Nucl.\ Phys.\ Proc.\ Suppl.\ }{\bf #1},}

\def\beq{\begin{equation}}
\def\eeq{\end{equation}}
\def\bea{\begin{eqnarray}}
\def\eea{\end{eqnarray}}
\newcommand{\ba}{\begin{array}}
\newcommand{\ea}{\end{array}}
\def\nn{\nonumber}

\title{Singlet extension of the MSSM as a solution to the small cosmological scale anomalies}

\author{Fei Wang$^1$, Wenyu Wang$^2$,  Jin Min Yang$^3$, Sihong Zhou$^2$}

\affiliation{$^1$ Department of Physics and Engineering, ZhengZhou University, ZhengZhou 450001, China\\
$^2$ Institute of Theoretical Physics, College of Applied Science,
              Beijing University of Technology, Beijing 100124, China\\
$^3$  State Key Laboratory of Theoretical Physics,
      Institute of Theoretical Physics, Academia Sinica,
              Beijing 100190, China}

\emailAdd{feiwang@zzu.edu.cn}
\emailAdd{wywang@mail.itp.ac.cn}
\emailAdd{jmyang@itp.ac.cn}
\emailAdd{sihongzhou@emails.bjut.edu.cn}

\abstract{In this work we show that the general singlet extension of the MSSM
can naturally provide a self-interacting singlino dark matter to
solve the small cosmological scale anomalies (a large Sommerfeld enhancement factor can also be obtained).
However, we find that the NMSSM (the singlet extension of the MSSM with $Z_3$ symmetry) cannot achieve
this due to the restricted parameter space.
In our analysis we introduce the concept of symmetric and antisymmetric viscosity cross sections
to deal with the non-relativistic Majorana-fermion dark matter scattering.}

\begin{document}
\maketitle \indent
\newpage

\section{Introduction}
As the standard model of the Big Bang cosmology, the $\Lambda$CDM model can account for most
observations of the Universe.  A crucial ingredient of this model is the existence of cold
dark matter (CDM), which, with a proper cosmological constant, can successfully predict the
large scale structure of the Universe.  However, the predictions on small scale structures
seem not so successful and some anomalies exist:~\cite{Bringmann:2013vra}
1) \textit{missing satellites} -- There should be many more
dwarf-sized subhalos (satellites) than observed in the DM halo of the Milky Way (MW).
And the observed galaxy luminosity and H{\sc i}-mass functions beyond the MW
show shallower faint-end slopes than predicted.
\cite{Klypin:1999uc}  
2) \textit{cusp vs ~core} -- It seems to have cored inner density profiles
in the low surface brightness and dwarf galaxies, this is at odds with CDM cusps predicted by
simulations~\cite{deNaray:2011hy}.
3) \textit{too big to fail} -- In comparison with the densest and most
massive satellites found in simulations, the observed brightest satellites of the MW attain their
maximum circular velocity at a too large radii.~\cite{BoylanKolchin:2011de}.

There are various ways to solve these small scale problems, such as the 
nonthermal production of warm dark matter \cite{Lin:2000qq} or the baryon 
feedback in the galaxies to make small halos dark \cite{Bullock:2010uy}.
Also, recently the author of \cite{Tulin:2012wi,Ko:2014nha} proposed another 
self-interacting Dirac-fermion DM scenario with a light mediator 
($\lsim$ 100 MeV) to solve these small scale anomalies.
With a light force carrier, the dark matter scattering cross section could have a non-trivial
velocity dependence. All of the small scales (the dwarf size, the Milky Way size as well as
the galaxy cluster size) can have appropriate cross sections, thus leaving enough parameter
space for the mass of DM, the force carrier and the coupling strength. Besides, the authors
also showed that the DM self-interactions can be correlated with the effect of Sommerfeld
enhancement in DM annihilation which is being probed through indirect detection experiments.

This self-interacting DM scenario perfectly explain the anomalies in
the simulations of small scale structures. Therefore, it is necessary to
check if such a scenario can be realized in popular new physics theories
like low energy supersymmetry (SUSY).
In SUSY the better known DM candidate is the Majorana-type neutralino, which is composed
of bino, wino and higgsinos. Apparently, if the neutralino can have self-interactions
through a light force carrier, such a carrier can not have sizable standard model (SM)
interaction due to the stringent constraints from both collider and DM detection experiments.
Thus, this light force carrier should be composed mainly of a singlet with respect to the
SM gauge groups, which cannot be found in the  minimal supersymmetric standard model (MSSM).

Fortunately, there are various singlet extensions of the MSSM, among which the
next-to-minimal supersymmetric standard model (NMSSM) seems most attractive \cite{fayet,NMSSM}.
In the NMSSM, all the parameters in the superpotential are dimensionless and electroweak
symmetry breaking is triggered by the TeV-scale soft SUSY breaking terms. The SUSY preserving
$\mu$ term in the superpotential of the MSSM is generated by the vacuum expectation values
(VEV) of a singlet superfield $S$.
It is shown that in the NMSSM a light singlet scalar at several GeV can survive the
DM detection limits and the collider constraints \cite{NMSSM2}.
On the other hand, if we do not impose any discrete symmetry 
(in the NMSSM it is $Z_3$) and allow for all possible interactions
of the singlet field, then we have the general singlet extension of the MSSM (GMSSM), 
(more detail can be seen in \cite{Kaminska:2014wia}) which was used
to explain the PAMELA anomaly \cite{Hooper:2009gm,Wang:2009rj}.
Compared with the NMSSM, the GMSSM has a larger parameter space.
In the GMSSM, the singlet can form a dark sector in case of a very small
$\lambda$. The singlino-like dark matter can annihilate 
into the light singlet-like scalar, which
can give the correct DM relic density and a proper Sommerfeld enhancement factor.
In this model, the singlet scalar can be even lighter than in 
the NMSSM due to a  larger parameter space.
So it is intriguing to check if such a singlet scalar in the NMSSM or GMSSM
can serve as the light force carrier
mediated in the DM self-interactions, which is the aim of this work.

In this work we focus on the NMSSM and GMSSM to check if the self-interacting DM scenario
can be realized. In our study we will take into account the constraints from DM relic density,
the DM direct detection experiments as well as the proper non-relativistic scattering cross sections
between DM.
We organize the content as follows.
In Sec. \ref{sec2}, we will discuss the general DM interactions.
In Sec. \ref{sec3} and Sec. \ref{sec4}, we will respectively
check the NMSSM and GMSSM to figure out the possibility of realizing the self-interacting DM
scenario to solve the small cosmological scale anomalies.
Sec. \ref{sec5} contains our conclusions.

\section{Dark matter interactions}\label{sec2}
\begin{figure}
\begin{center}\scalebox{1}{\epsfig{file=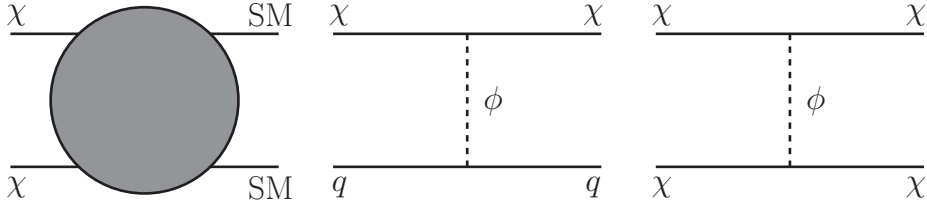}}\end{center}
\vspace{-0.7cm}
\caption{Dark matter interactions: annihilation to SM particles in the left diagram,
scattering off quarks in the middle diagram, and self-scattering in the right diagram.}
\label{fig1}
\end{figure}

 As mentioned in the introduction, in order to explain both the 
large scale and small scale structures of the Universe,
we can introduce the self-interacting DM scenario. The interactions between DM and SM particles
can be summarized as (shown in Fig. \ref{fig1}):
\begin{enumerate}
\item The annihilation to the SM particles (the left diagram of Fig. \ref{fig1}),
      whose cross section at high energy determines the relic density of dark matter
      and whose cross section at low energy is being probed by the indirect detection
       experiments like
       PAMELA \cite{pamela} and AMS02 \cite{AMS-2013}.
\item The elastic scattering off the SM particles (the middle diagram of Fig. \ref{fig1}),
      which is being probed by various direct detection experiments like
      CDMS,\cite{cdms2} XENON \cite{XENON100} and LUX \cite{lux}.
\item The non-relativistic self-scattering (the right diagram of Fig. \ref{fig1}),
      where $l=0$ in the partial wave expansion gives the Sommerfeld
       enhancement relative to the relativistic annihilation while $l\lsim 25$
       can account for the anomalies in the small cosmological scales.
\end{enumerate}
These interactions have subtle correlations among each other. A complete study on
DM properties needs to combine all these interactions from various  experiments.

We now briefly review the calculation of DM relic density and the DM-nucleon cross sections.
When the early universe was cooling down, the equilibrium between DM and SM particles
in the thermal bath can no longer be maintained. The DM will annihilate to SM particles
until the annihilation rate falls below the expansion rate of the universe. Thus the key
point in the DM relic density calculation is the annihilation rate of DM. In our following
calculation, we use the standard method \cite{susy-dm-review} to calculate the relativistic
annihilation cross section and the degrees of freedom at the freezing out temperature.

It is sufficient to consider only the spin independent (SI) elastic cross sections between
DM (denoted by $\chi$) and nucleon (denoted by $f_p$ for proton and $f_n$ for neutron
\cite{susy-dm-review} ) because of its high sensitivity in current DM direct detection
experiments. These interactions are dominantly induced by scalar exchange processes
at tree level, as shown in the middle diagram of Fig.(\ref{fig1}). Note that the vector
boson exchange interactions are also possible and we concentrate on the scalar interactions
only. For moderately light scalar bosons, $f_p$ is approximated by (similarly for $f_n$)
\begin{equation}
 \begin{split}
    f_{p}  \simeq
\sum_{q=u, d, s} \frac{f_q^{\phi}}{m_q} m_p f_{T_q}^{(p)}
    + \frac{2}{27}f_{T_G} \sum_{q=c, b, t} \frac{f_q^{\phi}}{m_q} m_p,
 \end{split}     \label{2b}
\end{equation}
where $f_{Tq}^{(p)}$ denotes the fraction of $m_p$ (proton mass)
from the light quark $q$, $f_{T_G}=1-\sum_{u,d,s}f_{T_q}^{(p)}$ is the heavy quark
contribution through gluon exchange, and $f_q^{\phi}$ is the
coefficient of the effective scalar operator given by
\begin{equation}
    f_q^{\phi} =  \frac{C_{\phi \chi  \chi}  C_{\phi qq}}{m_{\phi}^2},
    \label{Higgs-contr}
\end{equation}
with $C$ being the corresponding interaction vertex and
$m_\phi$ the mass of the exchanging particle.
The DM-nucleus scattering rate is then given by
\begin{equation}
    \sigma^{SI} = \frac{4}{\pi}
    \left( \frac{m_{ \chi} m_T}{m_{\chi} + m_T} \right)^2
    \times \bigl( n_p f_p + n_n f_n \bigr)^2,
\end{equation}
where $m_\chi$ is the DM mass, $m_T$ is the mass of target nucleus,
and $n_p (n_n)$ is the number of protons (neutrons) in the target nucleus.
We can see the dependence of SI cross section on the mass of exchanging
particle $\sigma^{SI}\propto 1/m_\phi^{4}$, which
is very important in our following discussions.

Note that the annihilation cross section in relic density
is calculated at high energy.
To explain the DM indirect detection results (such as the PAMELA result)
and the small cosmological scale anomalies,
the DM scattering at low energies are needed.
We will discuss their relations in the following.

\subsection{The Sommerfeld enhancement effect}
The Sommerfeld enhancement effect in dark matter annihilation is
proposed to explain some DM indirect detection results,
such as the positron excess observed by PAMELA \cite{pamela} or AMS \cite{AMS-2013}.
 The explanation of positron excess requires a very large DM annihilation rate
which on the other hand can not explain the DM relic abundance if the DM
is produced thermally in the early universe.
The Sommerfeld effect can greatly enhance the annihilation rate when the velocity
of DM is much smaller than the velocity at freeze-out temperature.
Note that it is shown that the positron excess can be more naturally explained
by the pulsar wind and, further, the gamma-rays
produced by inverse Compton scattering on interstellar radiation field
of electrons and positrons produced by dark matter basically exclude
the dark matter interpretation \cite{Cholis:2013psa}.

The Sommerfeld enhancement is a common effect in non-relativistic quantum mechanics (QM).
When DM has self-interactions, it will lead to an effective potential $V(r)$
in the non-relativistic limit. Thus the Schr\"{o}dinger equation of DM particle
can be written as
\begin{equation}
  -\frac{1}{2m_\chi}\nabla^2\psi_k+V(r)\psi_k=\frac{k^2}{2m_\chi}\psi_k , \label{sde3}
\end{equation}
where $k$ is the relative momentum of the
DM particle.  A non-relativistic DM moves and annihilates
around the origin, namely near $r = 0$. Therefore, the only effect of the
potential $V(r)$ is the change of the modulus for the wave-function at the origin
comparing to that without $V(r)$. Then, the annihilation cross section is enhanced to
\begin{equation}
\sigma = \sigma_0 S_k,
\end{equation}
where the Sommerfeld enhancement factor $S$ is given by
\begin{equation}
S_k = \frac{|\psi_k(0)|^2}{|\psi^{(0)}_k(0)|^2} = |\psi_k(0)|^2
\end{equation}
with $\psi_k(0)$ and $\psi_k^{(0)}(0)$ are respectively  the perturbed (unperturbed)
wave function of DM.

We can find a solution for the Schr\"{o}dinger equation
in non-relativistic QM. As $R_{kl}(r)\sim r^l$ as $r\to 0$ and the $l>0$ part
will not change the annihilating
wave function at the origin, we need only to calculate the $l=0$ part of the
radial wave $R_{kl}(r)$. We can numerically solve the Schr\"odinger equation
with the boundary condition
\begin{equation}
  \chi(r)\equiv rR_{kl}(r) \to \sin(kr+\delta) ~~{\rm as}~~r\to\infty.
\end{equation}
In case of a Yukawa-type effective potential induced by scalar exchange
\begin{equation}
  V(r)=-\alpha_\chi\frac{e^{-m_{\phi} r}}{r},
\end{equation}
with $\alpha_\chi={|C_{\phi\chi\chi}|^2}/{4\pi}$,
the Sommerfeld enhancement factor is given by
\begin{equation}
 S_k=\left|\frac{\frac{d\chi}{dr}(0)}{k}\right|^2.
\end{equation}

\begin{figure}[htb]
\begin{center}\scalebox{0.5}{\epsfig{file=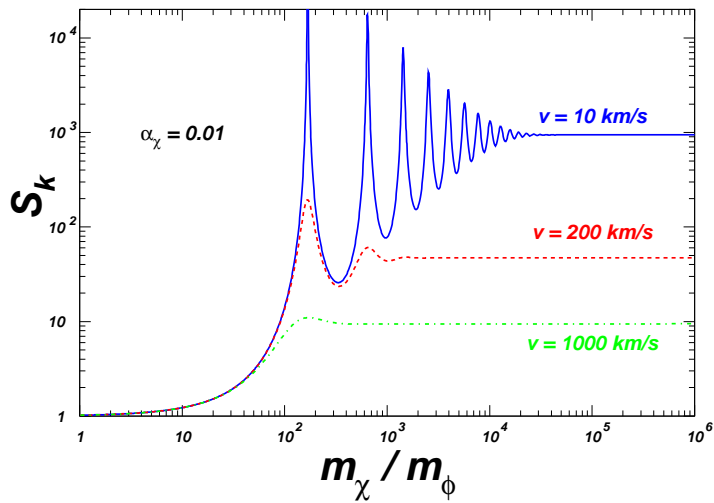}}\end{center}
\vspace{-0.7cm}
\caption{The Sommerfeld enhancement factor at different velocities with
the coupling strength $\alpha_\chi=0.01$. Here $v=10\rm km/s, 200\rm km/s,1000\rm km/s$
correspond to the characteristic speed of dwarf halos, the Milky Way and clusters,
respectively.}
\label{fig2}
\end{figure}
Though there is no analytical solution for the Schr\"{o}dinger equation with
the Yukawa-type potential,
there are three distinguishable regions for the Sommerfeld enhancement,
depending on the value of $m_\chi/m_\phi$. If the mediator scalar mass is comparable to
the DM mass, $S_k$ is negligible at all scales.
However, if the mediator scalar mass is much smaller than the DM mass,
the enhancement factor becomes independent of $m_\chi/m_\phi$ and corresponds to
the Coulomb limit. In this limit, the Sommerfeld enhancement factor is essentially
given by $S_k\sim\pi\alpha_\chi/v$.
In the resonance regions where $m_\chi/m_\phi\simeq\pi^2n^2/6\alpha_\chi$ with $n=1,2,3,\cdots$,
the DM annihilation cross section can be enhanced. The enhancement factor is approximately
given by $S_k\sim\pi^2\alpha_\chi m_\phi/(6m_\chi v^2)$, which is very sensitive to the DM velocity.
In our numerical calculation we reproduced the results of \cite{Tulin:2013teo}, as shown in
Fig. \ref{fig2}. This figure shows the Sommerfeld enhancement factor at velocity
$10,~200,~1000$ km/s, corresponding to the characteristic velocities of the halo of
the dwarf, the Milky Way and the cluster (these are the three small cosmological scales
at which the $\Lambda$CDM model seems not work well).

\subsection{The transfer and viscosity cross sections of self-interacting dark matter}
As pointed in \cite{Tulin:2013teo,Laha:2013gva}, in order to solve the small scale simulation anomalies,
self-interaction between the DM is necessary.
In the non-relativistic limit, the scattering between DM can be described by QM.
The most recent simulations have shown that $\sigma/m_\chi \sim 0.1 - 10 \, \cmg$ on dwarf scales
(the characteristic velocity is 10 km/s) is sufficient to solve the \textit{core-vs-cusp} and \textit{too-big-to-fail} problems,
while the Milky Way (the characteristic velocity is 200 km/s)
and cluster scales (the characteristic velocity is 1000 km/s) require
$\sigma/m_\chi \sim 0.1 - 1 \, \cmg$.  It appears that all the data may be
accounted for with a constant scattering cross section around
$\sigma/m_\chi \sim 0.5 \, \cmg$.
 On the other hand, the self-interacting DM models generically predict a velocity-dependent
scattering cross section over a wide range of parameter space.

 The numerical input for the simulation of small scales is the differential cross section
$d \sigma/ d \Omega$ as a function of the DM relative velocity $v$.
The standard cross section $\sigma = \int d\Omega (d\sigma/d\Omega)$ receives a strong
enhancement in the forward-scattering limit
($\cos\theta \to 1$), which does not change the DM particle trajectories.
Thus two additional cross sections are defined to parameterize
transport~\cite{krstic:1999}, the transfer cross section $\sigma_T$ and the viscosity
 (or conductivity) cross section $\sigma_V$:
\beq
\sigma_T =  \int d\Omega \, (1-\cos\theta) \,\frac{d\sigma}{d\Omega}  \, , \qquad  \label{sigmaT}
\sigma_V =  \int d \Omega \, \sin^2 \theta \, \frac{d\sigma}{d\Omega} \, .
\eeq
The transfer cross section is weighted by $(1-\cos\theta)$ which is the fractional longitudinal
momentum transfer while the viscosity cross section is weighted by $\sin^2 \theta$ which is the
energy transfer in the transverse direction.  The transfer cross section was used in the DM
literature to regulate the forward-scattering divergence in case of Dirac-fermion
DM candidate, while the viscosity cross section was used in case of Majorana-fermion DM
candidate. This is because both forward and backward scatterings diverge,
corresponding to poles in the $t$- and $u$-channel diagrams for the identical DM candidate.

In \cite{Tulin:2013teo} the authors considered a Dirac-fermion DM and calculated
$\sigma_T$. Within the resonance region, no analytic formula exists for $\sigma_T$
 and it must be computed by solving the Schr\"{o}dinger equation directly with partial
 wave expansion method. The scattering amplitude is given by
\beq \label{diffamp}
f(\theta) = \frac{1}{k} \sum_{\ell = 0}^\infty (2 \ell + 1)
e^{i \delta_\ell} P_\ell(\cos\theta) \sin \delta_\ell  \, .
\eeq
 The differential scattering cross section is given by
\beq \label{diffsigma}
\frac{d \sigma}{d\Omega} = \frac{1}{k^2} \Big| \sum_{\ell = 0}^\infty (2 \ell + 1)
e^{i \delta_\ell} P_\ell(\cos\theta) \sin \delta_\ell \Big|^2 \, ,
\eeq
where $\delta_\ell$ is the phase shift for a partial wave $\ell$.
 In terms of the phase shifts, the transfer cross section is given by
\beq
\frac{\sigma_T k^2}{4\pi} =  \sum_{\ell = 0}^{\infty} (\ell + 1)
\sin^2 (\delta_{\ell+1} - \delta_\ell)~.
\label{sigmaTsum}
\eeq
To obtain $\delta_\ell$, one must solve the Schr\"{o}dinger equation
to calculate the radial wave function $R_\ell(r)$ for the reduced
two-DM particle system
\beq
\frac{1}{r^2} \frac{d}{dr} \Big( r^2 \frac{d R_{\ell}}{dr} \Big)
+ \Big( k^2 - \frac{\ell (\ell + 1)}{r^2} - 2\mu V(r) \Big) R_\ell = 0. \label{radial}
\eeq
The Schr\"{o}dinger equation can be recast into the form
\beq \label{radial2}
\left( \frac{d^2 }{d x^2} +  a^2 - \frac{\ell(\ell+1)}{x^2}
\pm \frac{1}{x} \, e^{-x/b} \right) \chi_\ell(x) = 0 \;
\eeq
with new variables
\beq
\chi_\ell \equiv r R_\ell \, , \quad x \equiv \alpha_\chi m_\chi r  \, , \quad
a \equiv \frac{v}{2\alpha_\chi} \, , \quad b \equiv \frac{\alpha_\chi m_\chi}{m_\phi} \; . \label{vardefs}
\eeq
\begin{figure}[htb]
\begin{center} \scalebox{0.75}{\epsfig{file=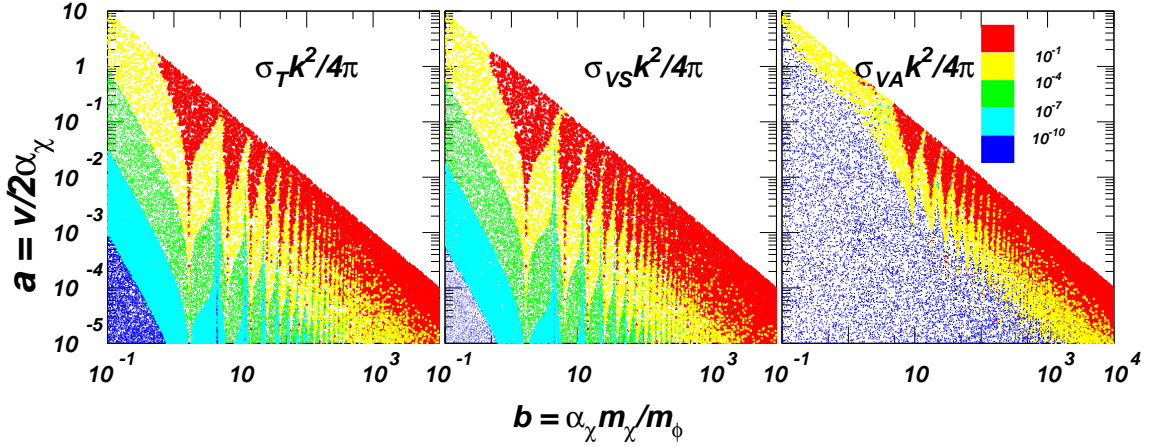}} \end{center}
\vspace{-0.8cm}
\caption{The birds-eye view of transfer and viscosity cross sections
$\sigma_T$, $\sigma_{VS}$, $\sigma_{VA}$ in the parameter space $(a, b)$.}
\label{fig3}
\end{figure}

When the DM candidate is a Majorana fermion, the amplitude and  Schr\"{o}dinger equation are
the same as in Eq.(\ref{diffamp}) and Eq.(\ref{radial2}). However, the total wave function
of the spin-1/2 fermionic DM must be antisymmetric with respect to the interchange of two
identical particles.
Then the spatial wave function should be symmetric when the total spin is 0 (singlet)
while the spatial wave function should be antisymmetric when the total spin is 1 (triplet).
The viscosity cross section should be defined with two variables:
\bea
\frac{d\sigma_{VS}}{d\Omega} &=& \left|f(\theta)+f(\pi-\theta)\right|^2 = \frac{1}{k^2}\left|
\sum_{\ell (\mbox{\tiny\rm EVEN\  number})}^{\infty} (2\ell + 1) (\exp(2i\delta_l)-1)P_\ell(\cos\theta)\right|^2
 \label{sigmaVSsum}\\
\frac{d\sigma_{VA}}{d\Omega} &=& \left|f(\theta)-f(\pi-\theta)\right|^2 = \frac{1}{k^2}\left|
\sum_{\ell (\mbox{\tiny\rm ODD\ number})}^{\infty} (2\ell + 1) (\exp(2i\delta_l)-1)P_\ell(\cos\theta)\right|^2
\label{sigmaVAsum}
\eea
Using the orthogonality relation for the Legendre polynomials,
we can obtain
\bea
\frac{\sigma_{VS}k^2}{4\pi} &=& \sum_{\ell(\mbox{\tiny EVEN\ number})}^{\infty}4\sin^{2}(\delta_{\ell+2}
-\delta_{\ell})(\ell+1)(\ell+2)/(2\ell+3) ,\label{sigmaVS}\\
\frac{\sigma_{VA}k^2}{4\pi} &=& \sum_{\ell(\mbox{\tiny ODD\ number})}^{\infty}4\sin^{2}(\delta_{\ell+2}
-\delta_{\ell})(\ell+1)(\ell+2)/(2\ell+3) . \label{sigmaVA1}
\eea
From the expressions of the transfer and viscosity cross sections,
we can see that both $\sigma_T$ and $\sigma_V$ will converge to a static value as
the phase shift $\delta_\ell$ approaching to a same value when the partial wave $\ell$ grows up.
We adopt the numerical method in \cite{Tulin:2013teo}
to calculate all the cross sections. The results are shown in Fig. \ref{fig3}
in which we have the same definitions of input variables.
The left plot is the reproduced results, same as in \cite{Tulin:2013teo}.
We can see that $\sigma_T$ and $\sigma_{VS}$ are almost same
while $\sigma_{VA}$ shows differences in some region ($ab \ll 0.1$).
The $ab > 0.5$ and $b\gg 1$ is the classical regime in which all the cross sections
can be very large.
There are also the resonance regions ($b \gtrsim 1$ and $ab \lesssim 0.5$) in which
all the cross sections exhibit patterns of resonance, making the cross section
much complicated. Self-interacting DM scenario used in \cite{Tulin:2013teo}
and in this paper works around the resonance regions. In the region $ab \ll 0.1$,
$\sigma_{VA}$ is much smaller than the other two cross sections.  The reason is
that the summation of the phase shift in $\sigma_{VA}$ begins
at $\ell=1$ as shown in Eq. (\ref{sigmaVA1}), and in this region the dominant
phase shift is $\delta_0$ and other phase shifts are much smaller.
 That is why we always calculate $\delta_0$ for partial wave expansion in
QM. In our following analysis, we assume that the DM scatters
with random orientations, thus the triplet is three times as likely as the
singlet and the average cross section will be
\bea
\sigma_{V} =\frac{1}{4}\sigma_{VS}+\frac{3}{4} \sigma_{VA}\; .\label{sigmaVa;;}
\eea
This is simple but sufficient to estimate the viscosity
cross section for Majorana DM.

\section{Can self-interacting dark matter scenario be realized in the NMSSM ? }\label{sec3}
SUSY can not only give a
solution to the hierarchy problem but also provide a good dark matter candidate
and realize gauge coupling unification. Among the SUSY models, the MSSM
has been intensively studied. This model, however, has the little hierarchy problem
since the newly discovered 125 GeV Higgs boson requires
a heavy stop or a large trilinear coupling $A_t$.
Besides, the MSSM also suffers from the $\mu$-problem \cite{muew}.
It is remarkable that both the little hierarchy problem and the
 $\mu$-problem can be solved in the NMSSM \cite{fayet,NMSSM}, in which an additional gauge
singlet $S$ is introduced. In this model the $\mu$-term is dynamically generated through
the coupling $SH_uH_d$ after $S$ develops an electroweak scale VEV, while
the little hierarchy problem is solved through an additional tree-level contribution
to the Higgs mass. With the additional singlet, it might be possible for the NMSSM to give DM
a proper non-relativistic cross section by tuning the singlet mediator. In the following we
check this possibility.

\subsection{Dark matter and Higgs bosons in the NMSSM}
In the NMSSM the relevant superpotential containing $\hat{S}$ is given by
\begin{eqnarray}\label{sp-nm}
\lambda\hat{S}\hat{H_u}\cdot\hat{H_d}+\frac{\kappa}{3}\hat{S}^3 \, ,
\end{eqnarray}
where $\hat{H}_u$ and $\hat{H}_d$ are the Higgs doublet superfields,
and $\lambda$ and $\kappa$ are dimensionless parameters.
Note that there is no explicit $\mu$-term and an effective $\mu$-parameter
is generated when the scalar component ($S$) of $\hat{S}$ develops a VEV $s$:
$ \mu_{eff}= \lambda s$.
The corresponding soft SUSY breaking terms are given by
\begin{eqnarray}\label{soft-term-nm}
A_\lambda \lambda S H_u\cdot H_d+\frac{A_\kappa}{3}\kappa S^3 +h.c.\, .
\end{eqnarray}
So the scalar Higgs potential is given by
\small
\begin{eqnarray}\label{pt}
V_F &=& |\lambda H_d\cdot H_u- \kappa S^2|^2
     + |\lambda S|^2 \left(|H_d|^2+|H_u|^2 \right)\, , \\
V_D &=&\frac{g_{2}^2}{2} \left( |H_d|^2|H_u|^2-|H_d \cdot H_u|^2 \right)
     +\frac{g_1^2+g_2^2}{8} \left( |H_d|^2-|H_u|^2\right)^2\, , \\
V_{\rm soft}&=&m_{d}^{2}|H_d|^2 + m_{u}^{2}|H_u|^2
            + m_s^{2}|S|^2  - \left( A_\lambda \lambda S H_d\cdot H_u
           + \frac{\kappa}{3} A_{\kappa} S^3 + h.c. \right)\, ,
\end{eqnarray}
\normalsize
where $g_1$ and $g_2$ are the coupling constant of $U_Y(1)$ and $SU_L(2)$, respectively.
Assuming
\beq
H_u^0 = h_u + \frac{H_{uR} + iH_{uI}}{\sqrt{2}} , \q
H_d^0 = h_d + \frac{H_{dR} + iH_{dI}}{\sqrt{2}} , \q
S = s + \frac{S_R + iS_I}{\sqrt{2}}
\eeq
with $h_u$, $h_d$, $s$ being the corresponding VEVs and using the minimization conditions,
one can obtain a $3\times3$ CP-even Higgs matrix ${\cal M}_h$,
a $3\times3$ CP-odd Higgs  matrix ${\cal M}_a$ and a $2\times 2$ charged Higgs matrix ${\cal M}_c$.
Note that there are three Goldstone bosons in ${\cal M}_a$ and ${\cal M}_c$ which
imply that after diagonalization these two matrices must have 0 eigenvalues.

From the superpotential in Eq. (\ref{sp-nm}), we can see that the interactions between the singlet
and the SM sector are controlled by the parameter $\l$.
If a light singlet Higgs exists, the constraints from both
collider and DM detections can be satisfied only if $\l$ is small enough.
Then the singlet Higgs will be a dark sector.
The spectrum of the NMSSM has been widely studied in the literature \cite{NMSSM},
so we only present our conventions and list two necessary terms for our analysis,
concentrating on the dark singlet sector:
\begin{itemize}
\item The CP-even Higgs mass matrix ${\cal M}_h$ in the interaction basis
$S^{bare} = (H_{uR}, H_{dR}, S_R)$ can be diagonalized by an orthogonal matrix $S_{ij}$ to
obtain 3 CP-even mass eigenstates (ordered in mass) $h_i = S_{ij} S^{bare}_j$
with the corresponding masses denoted by $m_{h_i}$. The elements involve the singlet
component are
\bea
{\cal M}_{h,33} & = & \l A_\l \frac{h_u h_d}{s}+ \k s (A_\k + 4 \k s),\\
{\cal M}_{h,13} & = & 2\l \mff h_u  - \l h_d (A_\l + 2\k s),\label{mh1}\\
{\cal M}_{h,23} & = & 2\l \mff h_d - \l h_u (A_\l + 2\k s) .
\eea

\item Neutralino matrix ${\cal M}_0$ is composed of
the $U(1)_Y$ gaugino $\l_1$, the neutral $SU(2)$ gaugino
$\l_2$, the singlino $\psi_s$ and the neutral higgsinos $\psi_{u,d}^0$.
In the basis $\psi^0 = (-i\l_1 , -i\l_2, \psi_u^0, \psi_d^0,
\psi_s)$ one can rewrite
\beq
{\cal L} = - \half (\psi^0)^T {\cal M}_0 (\psi^0) + \mathrm{h.c.}
\eeq
where
\beq
{\cal M}_0 =
\left( \ba{ccccc}
M_1 & 0 & \frac{g_1 h_u}{\sqrt{2}} & -\frac{g_1 h_d}{\sqrt{2}} & 0 \\
0& M_2 & -\frac{g_2 h_u}{\sqrt{2}} & \frac{g_2 h_d}{\sqrt{2}} & 0 \\
\frac{g_1 h_u}{\sqrt{2}} & -\frac{g_2 h_u}{\sqrt{2}} & 0 & -\mff & -\l h_d \\
-\frac{g_1 h_d}{\sqrt{2}}&\frac{g_2 h_d}{\sqrt{2}} &-\mff & 0 & -\l h_u \\
0& 0&  -\l h_d&-\l h_u & 2 \k s
\ea \right) . \label{neutralino_matrix}
\eeq
After diagolization, one obtains 5 eigenstates (ordered in mass) $\chi^0_i = N_{ij} \psi^0_j$,
among which the lightest one  $\chi^0_1$  is usually assumed to be the lightest
SUSY particle and serve as a good DM candidate.
\end{itemize}
From the spectrum we can see that if $\l$ approaches zero,
one can tune the masses of the singlet-dominant scalars and neutralinos
to arbitrary values by varying the parameters $\k,~A_\k$.

DM (the lightest neutralino) in the NMSSM has three-type components: gaugino, higgsino and singlino.
Assuming the gaugino unification relation $M_2/M_1\approx 2$, we have  three possibilities
for the DM:
\begin{itemize}
\item Bino-dominant DM. As shown in \cite{Cao:2010ph}, under current collider
and DM relic density constraints, the SI cross section can exclude a large part
of parameter space, leaving only a bino-dominant DM candidate below TeV.
\item Higgsino-dominant DM. As pointed in \cite{Wang:2013rba}, the higgsino-dominant DM
candidate around 1.1 TeV can satisfy all the constraints, including the relic density
and current DM direct detections.
\item Singlet-dominant DM. In order to explain the observation of CoGeNT \cite{Aalseth:2010vx},
the analysis in \cite{Draper:2010ew} showed that in the Peccei-Quinn limit there can
exist three light singlet-like particles (0.1-10 GeV): a scalar,
a pseudoscalar and a singlino-like DM candidate.
For a certain parameter window, through annihilation into the light pseudoscalar
the singlino DM can give the correct relic
density, and through exchanging the light scalar in scattering off the nucleon
a large cross section suggested by CoGeNT and DAMA/LIBRA \cite{Bernabei:2010mq}
 can be attained.
\end{itemize}

\subsection{Allowed parameter space and dark matter self scattering in the NMSSM}
We use the package NMSSMTools \cite{nmssmtools} to study all the scenarios numerically. As the
package requires the mass of light Higgs to be heavier than 1 GeV, we modify the package so that
the light Higgs mass can be arbitrarily low and also we include our own codes in the
calculation of SI cross section. Since it is nontrivial to obtain a light singlet-dominant Higgs,
we randomly scan the parameter space under the condition
$m_{h_1} < 5 {\rm GeV}$ and $m_{h_2} > 120 {\rm GeV}$.
In our scan we consider constraints from DM relic density and DM-nucleon scattering
cross section.
Since the SM Higgs ($h_2$) is much heavier than the singlet-like Higgs ($h_1$), we only
need to slightly tune the parameter $A_\k$ to get a light Higgs with mass 100 MeV.
Finally, we use a light scalar with $m_{h_1} = 100 {\rm MeV}$ for the calculation of
 DM-nucleon SI cross section $\sigma^{\rm SI}$.
Note that the singlet sector will be dark with respect to  the SM sector when $\l$ is
sufficiently small, and then the SM sector is very insensitive to the tuning
in the dark sector.

\begin{figure}[htbp]
\begin{center}\scalebox{0.5}{\epsfig{file=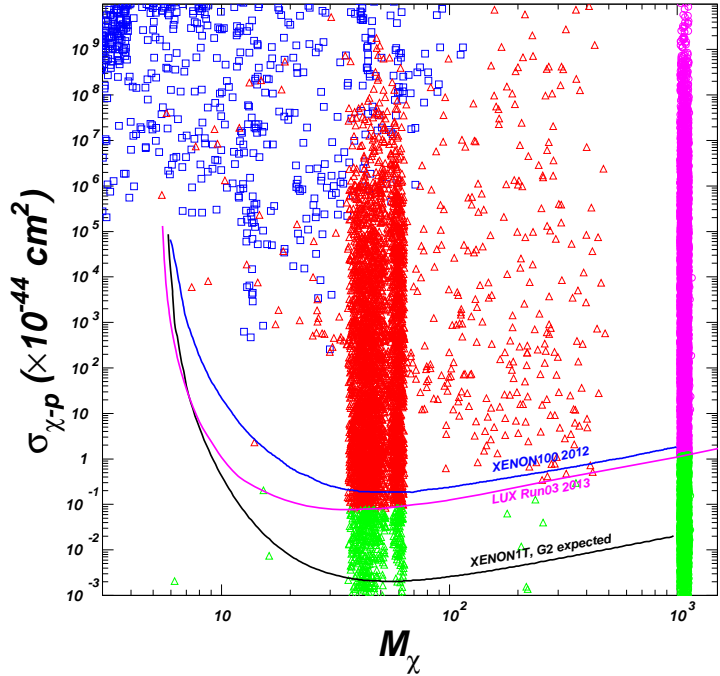}}\end{center}
\vspace{-0.7cm}
\caption{The spin-independent cross section of DM scattering off the proton
versus the DM mass
in different scenarios in the NMSSM. $\Box$ denotes singlino-dominant
DM scenario, most of which are excluded by the DM direct detection limits;
$\bigtriangleup$ denotes bino-dominant DM scenario, among which
the red ones are excluded while the green ones survive;
$\circ$ (around 1 TeV) denotes the higgsino-dominant DM scenario,
among which the pink ones are excluded while the green ones survive.
All the samples satisfy the DM relic density constraints. }
\label{fig4}
\end{figure}
The numerical results of the three possibilities are shown in Fig. \ref{fig4},
in which the SI cross section of DM-nucleon scattering is shown and current
constraints of XENON and LUX are displayed. We can see that the DM direct
detection constraints exclude most of the samples.
However, there are still some surviving parameter space, especially for
bino-dominant and higgsino-dominant scenarios. Future XENON1T will further
cover the surviving parameter space.

Now we present some details related to our scan for different scenarios:
\begin{itemize}
 \item For the singlino-dominant DM scenario,
  we define the parameter $\varepsilon\equiv {\lambda \mu}/{m_Z}$,
$\varepsilon' \equiv {A_\lambda}/{\mu\tan\beta} -1$ and scan the parameter space
as in \cite{Draper:2010ew}:
\bea
&& 2\leq\tan\beta\leq 50, ~0.05\leq\lambda\leq 0.2,
~0.0005\leq\kappa\leq0.05, -0.1\leq\varepsilon \leq 0.1,\nn\\
&& \varepsilon \sim \varepsilon', |A_\k|<500{\rm GeV}, |\mu|<1 {\rm TeV} .
\eea
The sfermion mass is set to be 6 TeV so that we can easily get a 125 GeV SM
Higgs. As noted in \cite{Draper:2010ew}, the DM annihilates to SM particles mainly
through the resonance of the singlet pseudo-scalar $a_1$ (the Feynman diagram is
shown in Fig. \ref{fig5}).
The result of the scan is shown in the left panel of Fig. \ref{fig6}.
As a consequence of resonance, the singlet pseudo-scalar mass is almost twice of the DM
mass. This result is consistent with \cite{Draper:2010ew} and also implies that the
coupling of $a_1f\bar f$ should not be completely neglected and thus the parameter
$\l$ can not be too small.
On the other hand, as shown in sec.\ref{sec2}, the SI cross section has the inverse
quartic power dependence on the mediated scalar mass (here it is the singlet Higgs $h_1$).
Though the coupling of $h_1f\bar f$ is suppressed by a small $\l$,
the cross section will be enhanced greatly when the mass of $h_1$ is below GeV.
Therefore, this scenario can hardly satisfy the direct detection limits on
the SI cross section.
\begin{figure}[htbp]
\begin{center}\scalebox{0.5}{\epsfig{file=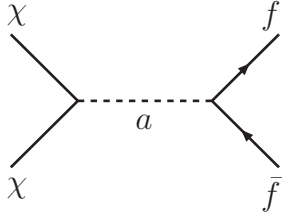}}\end{center}
\vspace{-0.7cm}
\caption{Feynman Diagram of DM annihilation to SM fermions through a pseudo-scalar $a$.}
\label{fig5}
\end{figure}

\begin{figure}[htbp]
\begin{center}\scalebox{0.6}{\epsfig{file=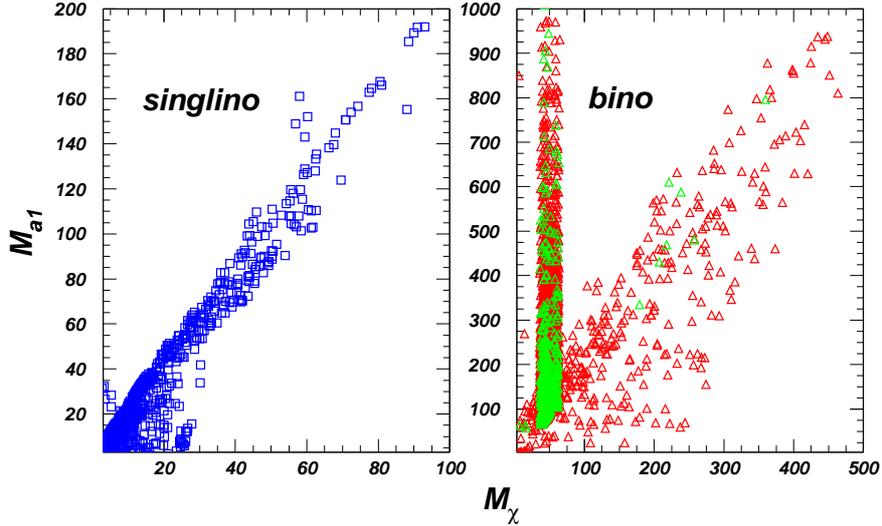}}\end{center}
\vspace{-0.7cm}
\caption{The scatter plots of parameter space allowed by DM relic density
shown in the plane of $M_{a_1}$ versus the DM mass.
The left panel is the singlino-dominant DM scenario, where most of the samples
are excluded points by direct detection limits of LUX-2013.
The right panel is the bino-dominant DM scenario, where
the red ones are excluded points by direct detection limits of LUX-2013
while the green ones survived.}
\label{fig6}
\end{figure}

\item For the bino-dominant DM scenario,
we scan the parameter space
\bea
 10^{-5}\leq|\l|,|\k|<1, ~~2\leq\tan\beta\leq 50,~~|\mu|,|A_\l|<1\rm TeV,
~~|M_2|<500{\rm GeV}~,
 \eea
and set $A_\k \sim -4\k\mu/\l$ to get a light singlet Higgs (see Eq. (\ref{mh1})).
We also keep the gaugino unification assumption $M_1=M_2/2, ~M_3=3M_2$ and set the soft
parameters to 6 TeV.

Since bino does not couple to singlet Higgs directly, the coupling between bino-dominant DM and
singlet Higgs arises from its higgsino and singlino components.
Therefore, the SI cross section of bino-dominant DM can be much less than the LUX limits,
as shown in Fig. \ref{fig4}.

The right panel of Fig. \ref{fig6} shows the samples that satisfy the DM relic density
in the bino-dominant DM scenario.
The samples can be divided into two regions: one is in the diagonal region($m_a\sim 2m_\chi$)
and the other is a vertical belt around $m_\chi\sim 50\rm GeV$.
The first region is excluded by the direct detection limits of LUX
(the reason is similar to the singlino-dominant DM scenario).
In the second region, the DM annihilates to SM fermions through the $Z$-boson
resonance, independent of singlet coupling $\l$. The coupling $\l$ is constrained
only by the SI cross section limits and can only take a very small value, namely
$\l\lsim0.005$. This can be seen from the left panel of Fig. \ref{fig7}.

\begin{figure}[htbp]
\begin{center}\scalebox{0.6}{\epsfig{file=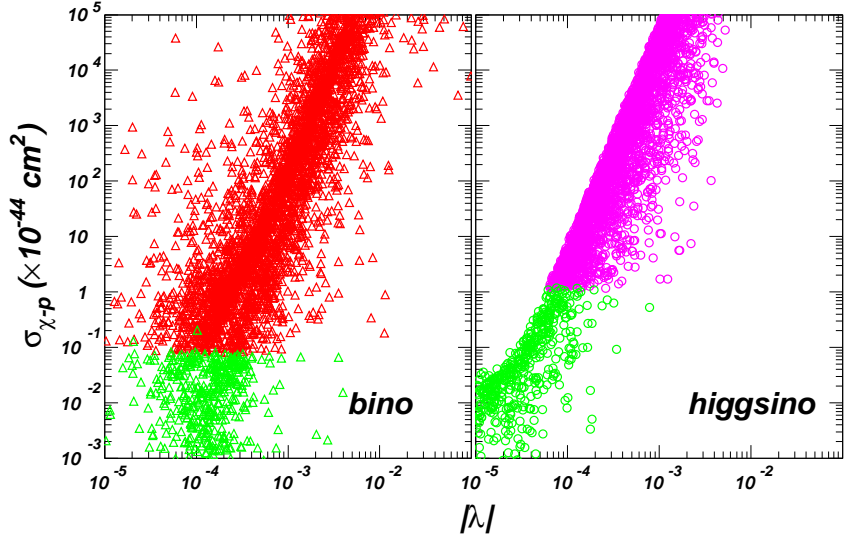}}\end{center}
\vspace{-0.7cm}
\caption{Same as Fig.\ref{fig4}, but showing the spin-independent cross section versus $\l$
in the NMSSM.}
\label{fig7}
\end{figure}

\item For the higgsino-dominant DM scenario,
 we scan the parameter space
\bea
&& 10^{-5}\leq|\l|,|\k|<1, ~~2\leq\tan\beta\leq 50, ~~|A_\l|<1\rm TeV, \nn \\
&& |\mu|<1.5\rm TeV, ~~A_\k \sim -4\k\mu/\l,~~|M_2|=10{\rm TeV}~.
\eea
As above, the gaugino unification assumption is used and the soft parameters are set to be 6 TeV.
In this scenario, the DM mass need to be around 1100 GeV to satisfy the relic density, and thus
no resonance through $Z$-boson or pseudo-scalar happens in the DM annihilation.

The singlet Higgs can couple directly to the higgsino through the $\l S\psi_u\psi_d$ term
in the Lagrangian, so the SI cross section limits only constrain $\l$. From the right panel
of Fig. \ref{fig7} we can see that $\l\lsim 0.001$ for the survived samples.
\end{itemize}

As a brief summary, the singlino-dominant DM scenario is almost excluded because of the
$\l$-value correlation between relic density and DM-nucleon SI cross section
(the relic density needs a not-too-small $\l$ which is too large for the DM-nucleon
SI cross section).
On the other hand, some parameter space still survived the SI cross section limits
in the bino-dominant and higgsino-dominant DM scenarios, in which
the $\l$-value correlation between relic density and DM-nucleon SI cross section is relaxed
and $\l$ is constrained by the DM-nucleon SI cross section limits to take a very small value.

\begin{figure}[htbp]
\begin{center}\scalebox{0.4}{\epsfig{file=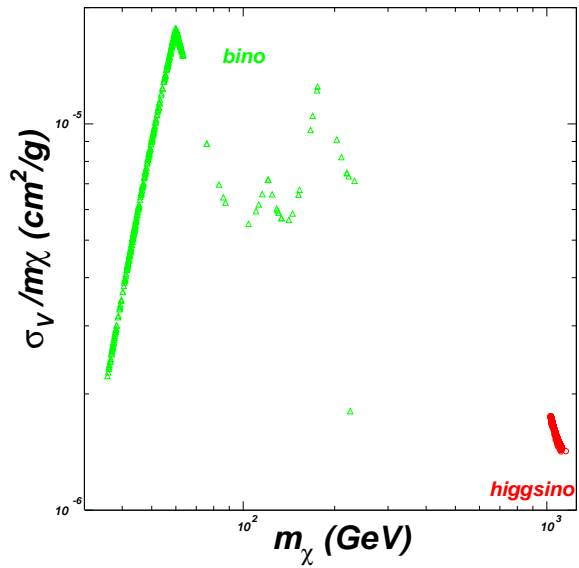}}\end{center}
\vspace{-0.7cm}
\caption{Same as Fig.\ref{fig7}, but showing the non-relativistic viscosity
cross section between DM in the NMSSM.
For each allowed sample in Fig. \ref{fig7}, we use three
velocities corresponding to the three small cosmological scales (dwarf, Milky Way and cluster)
to calculate the cross section and here we plot the maximal one.}
\label{fig8}
\end{figure}

In order to figure out if the NMSSM can realize the DM self-interaction scenario to
explain the small cosmological scale anomalies, we now calculate the DM self scattering
cross section. For each allowed sample in Fig. \ref{fig7}, we use three
velocities corresponding to the three small cosmological scales (dwarf, Milky Way and cluster)
to calculate the DM self scattering cross section. Then for each sample we have three
values of  DM self scattering cross section and we keep the maximal one to display it
in Fig. \ref{fig8}. We see that due to a very small $\l$ value, $\sigma_V/m_\chi$ is too small
to explain the small cosmological scale anomalies
(its value must be in $0.1 - 10 \, \cmg$ to explain dwarf scale anomaly
and in $0.1 - 1 \, \cmg$ to explain MW and cluster anomalies).
For such a weak scattering, we can regard the DM as a collisionless particle.
So we conclude that NMSSM can not realize the self-interacting DM scenario
to solve the small cosmological scale anomalies.

\section{Can self-interacting dark matter scenario be realized in the GMSSM ?}
\label{sec4}
The main reason for the failure of the NMSSM in solving the small cosmological
scale anomalies is a too small $\l$ (constrained by DM-nucleon SI cross section
limits), which determines the coupling strength of $h_1\chi\chi$ in DM
self-interaction (Note that when the DM is singlino-dominant, the coupling
$h_1\chi\chi$ can be proportional to $\k$. However, a singlino-dominant DM
is obtained under the condition $\k \ll \l$ \cite{Draper:2010ew}, which
means a very small $\k$).
In the NMSSM the $\l S H_u \cdot H_d$ term
in the superpotential is the origin of the $\mu$-term.
In the GMSSM, however, the ${\cal Z}_3$ discrete symmetry is not imposed and
the $\mu$ term can exist in the superpotential, together with
the $\l S H_u \cdot H_d$ term (several other terms
of the singlet superfield can also exist in  the superpotential).
Then the singlet sector can be a completely dark
sector in case of a nearly vanishing $\l$. Unlike the NMSSM,
the dark Higgs sector (including a singlino-dominant DM)
can be easily realized in the GMSSM, which does not need the condition
 $\k \ll \l$ \cite{Draper:2010ew}. This means that a singlino-dominant DM can be
obtained with a sizable  $\k$ and in this case the coupling
$h_1\chi\chi$ in DM self-interaction, which is proportional to $\k$,
can be large. So it should be easy to realize the self-interacting DM
scenario in GMSSM.

As shown in \cite{Wang:2009rj}, in the GMSSM a singlino-dominant dark matter
annihilates to a singlet-dominant Higgs, which can give the correct DM relic density,
and also the DM-nucleon SI cross section limits from the direct
detections can be easily satisfied.
In addition, an appropriate Sommerfeld enhancement can explain the positron excess
observed by PAMELA.
The non-relativistic DM scattering and Sommerfeld enhancement are similar to
what discussed in \cite{Tulin:2012wi} except that there can mediated an
additional pseudo-scalar in the GMSSM.

\subsection{Dark matter and Higgs bosons in the GMSSM}
 The superpotential of the GMSSM is
\bea
W = \mu \widehat{H}_u \cdot \widehat{H}_d+\l \widehat{S}
\widehat{H}_u \cdot \widehat{H}_d+\eta \widehat{S}
+\half \mu_s \widehat{S}^2 + \frac{1}{3} \k \widehat{S}^3 \ ,\label{sp-gm}
\eea
which involves the parameters $\mu,~\l,~\eta,~\mu_s,~\k$.
We set $\l\sim 0$ so that the singlet sector almost decouple from the SM sector.
In the following discussion, we will concentrate on the singlet sector.

The soft SUSY breaking terms take the form
 \bea
 -{\cal L}_\mathrm{soft} &=&  m_s^2 | S |^2
 + \left( C_\eta \eta S +\half B_s \mu_s S^2 + \third \k A_\kappa\ S^3 + \mathrm{h.c.}
 \right)\,. \label{soft-gm}\eea
The main difference between the NMSSM and GMSSM is reflected in their Higgs sectors
which contain different singlet Higgs mass matrices and self-interactions.
The difference mainly comes from the F-term $F_s$:
\bea
V_{F_s} &=& |F_s|^2=|\eta+\mu_s S +\k S^2|^2\nn\\
&=& |\k S^2|^2 + \eta^2+\mu_s^2|S|^2
+\left(\eta\mu_s S  +\k \eta S^2 +\k\mu_s S^2 S^\ast + \mathrm{h.c.}\right) .
\eea
Since $\eta^2$ is a constant, the $\mu_s^2|S|^2, \eta\mu_s S,\k \eta S^2$ terms
can be absorbed by the redefinition of the soft SUSY breaking parameters
$m_s^2|S|^2,~ C_\eta \eta S,\half B_s \mu_s S^2$.

The singlet Higgs potential is
\bea
V & = & V_F+V_{\rm soft}\nn\\
&=&   m_s^2|S|^2 + \left(C_\eta \eta S+\half B_s \mu_sS^2+ \third \k A_\k\ S^3 +\k\mu_s S^2 S^\ast + \mathrm{h.c.}\right).
\label{vform}
\eea
The singlet chiral supermultiplet contains a complex scalar and a
Majorana fermion $\chi$.
After the scalar component getting a VEV,  we can get one CP-even Higgs $h$ and
one CP-odd Higgs $a$. The mass spectrum and the relevant Feynman rules are
\bea
m_{\chi}&=&2\k s+\mu_{S}, \\
m_{h}^{2}&=&\k s(4\k s+A_{\k}+3\mu_{S})-C_\eta\eta /v, \\
m_{a}^{2}&=&-2B_{s}\mu_{s}-\k s(3A_{\k}+\mu_{S})-C_\eta\eta/s, \\
V_{hhh} &=& -4ik(6\k s+A_{\k}+3\mu_{S})=-4i\k(3m_{\chi}+A_{\k}),\\
V_{haa} &=& -4i\k(2\k s-A_{\k}+\mu_{S})=-4i\k(m_{\chi}-A_{\k}),\\
V_{h\chi\chi}&=&-4i\k,\\
V_{a\chi\chi}&=&-4\k\gamma^{5}.
\label{gm-fr}
\eea
We do not give the Feynman rules for the four-scalar vertex for they are irrelevant
to our analysis. From the spectrum and Feynman rules above, we can see that the
mass parameters of the three particles ($h, a,  \chi$)
can be set freely because the relevant input soft parameters
can take arbitrary values.
 This makes the following calculation much easier and we can choose
the input parameters easily without fine-tuning.

Then we have only five input parameters in our calculation,
namely $m_\chi,~m_h,~m_a,~\k,~A_\k$.
It is useful to note that the three CP-even Higgs vertex $V_{hhh}$
can vanish if we set $A_{\k}=-3m_{\chi}$.

\begin{figure}[htbp]
\begin{center}\scalebox{1}{\epsfig{file=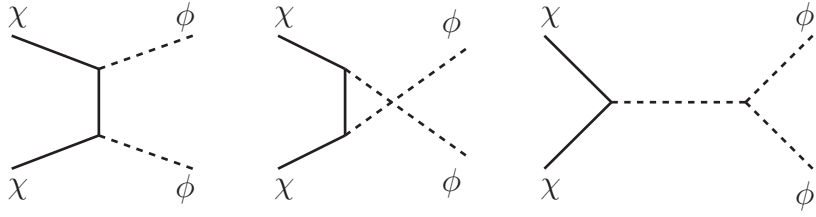}}\end{center}
\vspace{-0.7cm}
\caption{Feynman diagram of singlino DM annihilation to scalars in the GMSSM.
The final state of $\phi~\phi$ can be $h h$, $h a$ and $a a$. }
\label{fig9}
\end{figure}
The calculation of non-relativistic DM scattering cross section is already given
in Sec. \ref{sec2} with the coupling strength of DM to mediator
($\alpha_\chi =4\k^2/\pi$).
Here we just give some details relevant to the calculation of DM relic density.
The Feynman diagrams of the DM annihilation are shown in Fig. \ref{fig9}.
The general form of the annihilation cross section is given by \cite{Drees:1992am}
\beq
\sigma v =\frac{1}{4}\frac{\bar\beta_f}{8\pi sS}\left[
|A(^1S_0)|^2+\frac{1}{3}\left(|A(^3P_0)|^2+|A(^3P_1)|^2\right)+|A(^3P_2)|^2\right],
\eeq
where $S$ is the symmetry factor, $A(^1S_0),~A(^3P_0),~A(^3P_1),~A(^3P_2)$
are the contributions from different spin states of DM and $\beta_f$ is given by
\bea
\bar\beta_f=\sqrt{1-2(m_X^2+m_Y^2)/s + (m_X^2+m_Y^2)^2/s^2}~,
\eea
with $X,Y$ being the final states.
The amplitudes from different final states are given by
\begin{enumerate}
\item $\chi\chi\to h~h$ :
\bea
A(^3P_0) &=& 16\sqrt{6}v\k^2\left[\frac{R(3m_\chi+A_\k)}{4-R(m_h)^2+iG_h}
-2\frac{1+R(m_\chi)}{P_\chi}+\frac{4}{3}\frac{\bar\beta_f^2}{P_\chi^2}\right] ,\\
A(^3P_2) &=& -(128/\sqrt{3})v\k^2\bar\beta_f^2 /P_\chi^2~.
\eea
\item   $\chi\chi\to a~a$ :
\bea
A(^3P_0) &=& 16\sqrt{6}v\k^2\left[\frac{R(m_\chi-A_\k)}{4-R(m_h)^2+iG_h}
-2\frac{1-R(m_\chi)}{P_\chi}+\frac{4}{3}\frac{\bar\beta_f^2}{P_\chi^2}\right], \\
A(^3P_2) &=& -(128/\sqrt{3}) v\k^2\bar\beta_f^2 /P_\chi^2~.
\eea
\item   $\chi\chi\to h~a$ :
\bea
A(^3S_0) &=& -32\sqrt{2}\k^2\frac{R(m_\chi-A_\k)}{4-R(m_a)^2}\left(1+\frac{v^2}{8}\right)
\nn\\
&& +64\sqrt{2}\k^2\frac{R(m_\chi)}{P_\chi}\left[1+v^2\left(\frac{1}{8}
-\frac{1}{2P_\chi}+\frac{\bar\beta_f^2}{3P_\chi^2}\right)\right]\nn\\
&& +16\sqrt{2}\k^2 \left(R(m_a)^2-R(m_h)^2\right)\left[1+v^2\left(-\frac{1}{8}
-\frac{1}{2P_\chi}+\frac{\bar\beta_f^2}{3P_\chi^2}\right)\right], \\
A(^3P_1) &=& 64v\k^2 \bar\beta_f^2 /P_\chi^2~.
\eea
In the above formulas
\bea
R(m_X)=\frac{m_X}{m_\chi},~P_\chi=1+R(m_\chi)^2-\frac{1}{2}(R(m_X)^2+R(m_Y)^2),
~G_h=\frac{\Gamma_h m_h}{m_\chi^2}.
\eea
\end{enumerate}
Given the annihilation amplitudes, the relic density $\Omega h^2$ can be
calculated in the standard way.
From the expression of the DM relic density \cite{Tulin:2013teo}
\beq
\sigma v =\frac{3}{8}\frac{\pi\alpha_\chi^2}{m_\chi^2}v^2\sqrt{1-\frac{m_h^2}{m_\chi^2}}~,
\eeq
we can see that the ratio $\alpha_\chi/m_\chi$ is almost a constant when
$m_h\ll m_\chi$ in order to give the correct DM relic density.

\subsection{Allowed parameter space and dark matter self scattering in the GMSSM}
\begin{figure}[htbp]
\begin{center}\scalebox{0.5}{\epsfig{file=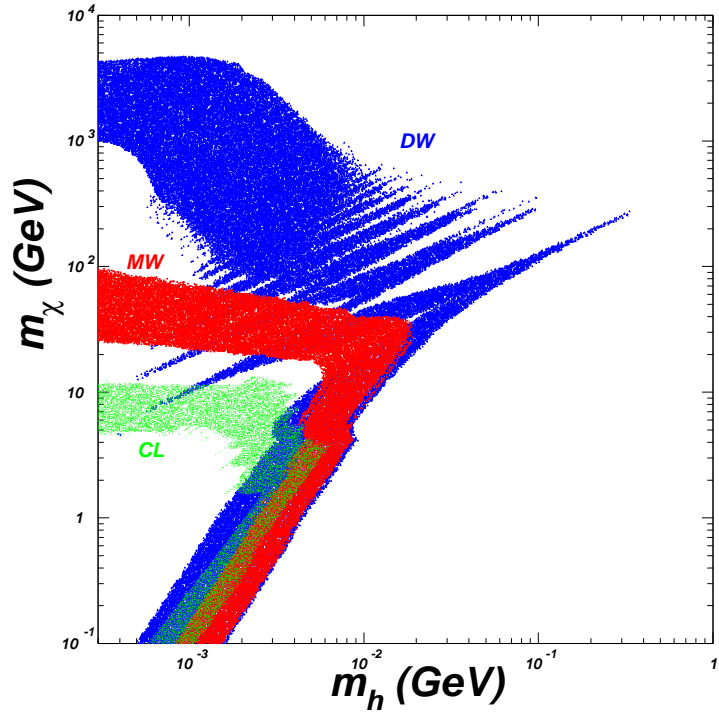}}\end{center}
\vspace{-0.7cm}
\caption{Scatter plots of the parameter space in the DM self-interaction
model \cite{Tulin:2013teo} allowed by DM relic density plus dwarf scale
structures (blue region),
Milky Way size structures (red region) or  cluster size structures
(green region). We obtained these results by changing our Feynman rules
to those in \cite{Tulin:2013teo}.}
\label{fig10}
\end{figure}

We first change our Feynman rules to those in \cite{Tulin:2013teo}
and check if we can reproduce the results in \cite{Tulin:2013teo}.
Our results are shown in Fig. \ref{fig10}, which agree with \cite{Tulin:2013teo}.
Fig. \ref{fig10} shows the transfer cross section
$\sigma_T$ allowed by solving the small cosmological scale anomalies
(dwarf, Milky Way and cluster) with their corresponding characteristic velocities.
Here all the samples satisfy the DM relic density given by PLANCK \cite{planck}.
We can see that $\sigma_T/m_\chi$ can satisfy simultaneously the requirements of
dwarf, Milky Way and galaxy cluster scales to solve the small scale anomalies.
On the other hand, as can be seen from the figure, the requirement to solve
such small scale structure anomalies can also give strong constraints on the
parameter space of self-interacting DM scenario. This means that subtle
relations should be satisfied between $m_h,m_\chi$ and $\alpha_\chi$.

\begin{figure}[htbp]
\begin{center}
\scalebox{0.4}{\epsfig{file=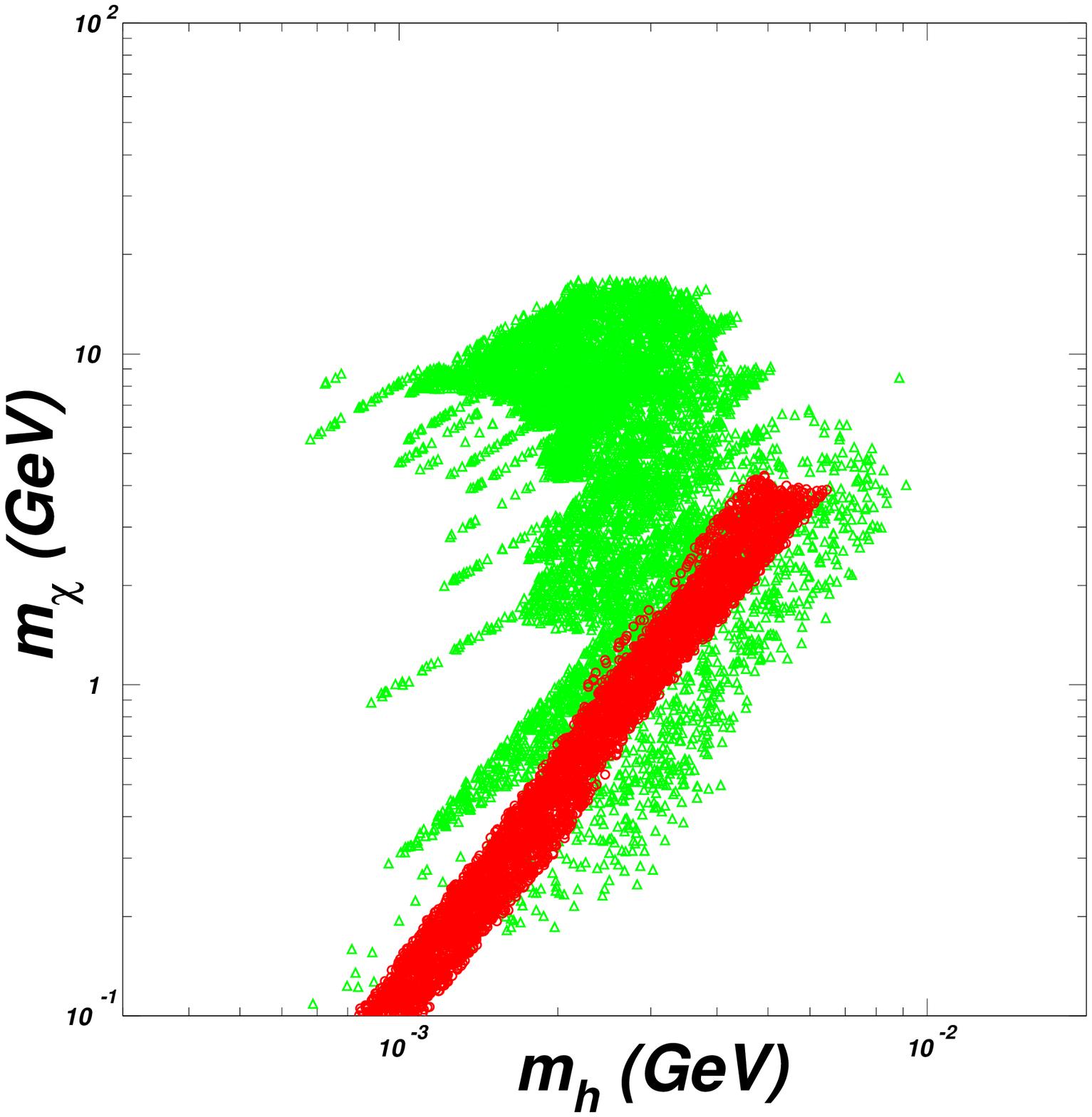}}
\scalebox{0.395}{\epsfig{file=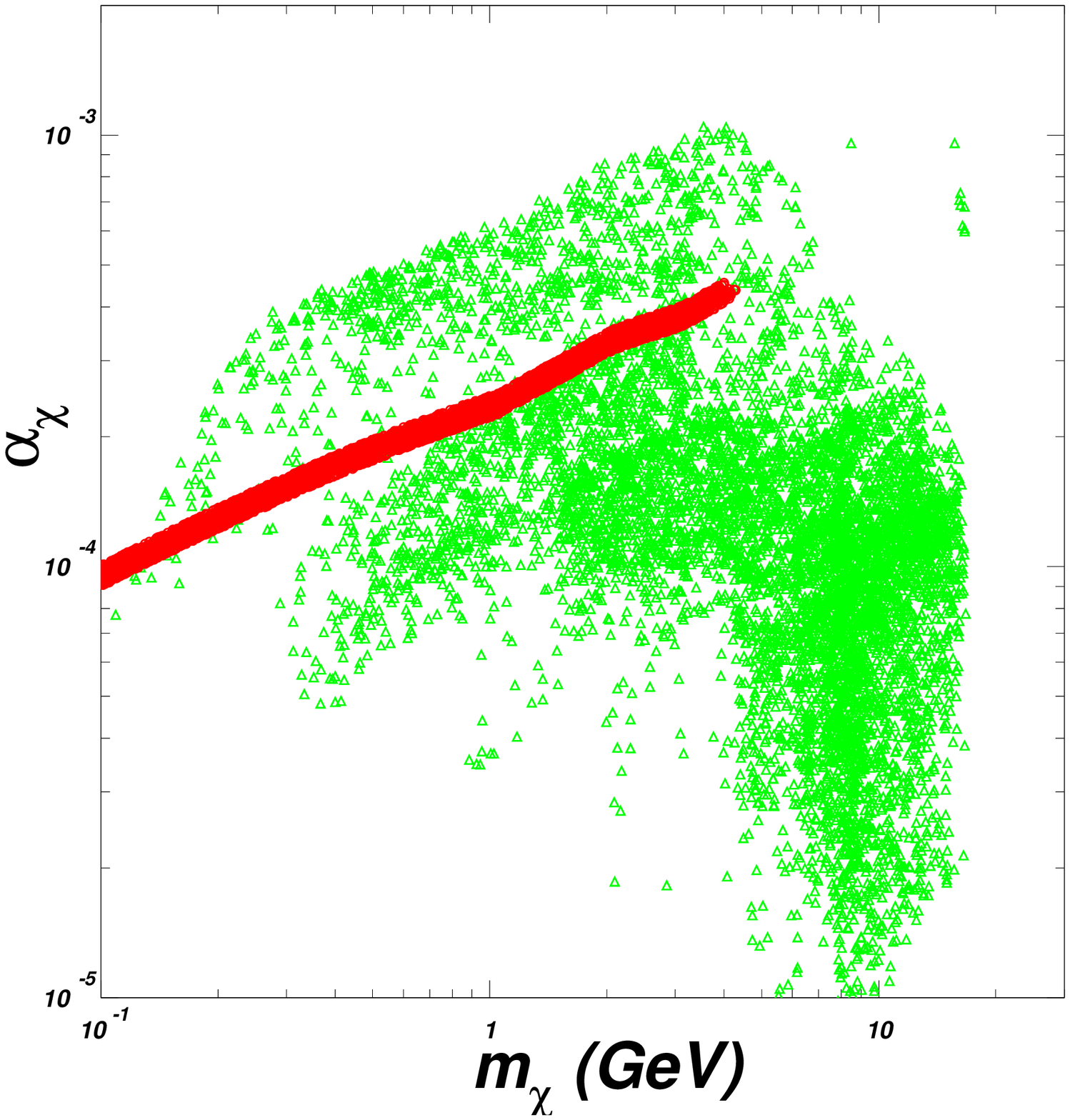,}}
\end{center}
\vspace{-0.7cm}
\caption{The scatter plots of the parameter space allowed by
DM relic density and all three small scale structures (dwarf,
Milky Way, cluster).
The red ones are for the DM self-interaction model \cite{Tulin:2013teo}
while the green ones are for the GMSSM.}
\label{fig11}
\end{figure}

Then we perform a numerical calculation in the GMSSM.
Since in the GMSSM the singlino DM is of Majorana type,
we need to use viscosity cross section $\sigma_V$.
In Fig. \ref{fig11} we show
the GMSSM parameter space allowed by
DM relic density plus all three small scale structures.
The corresponding parameter space for the DM self-interaction model \cite{Tulin:2013teo}
is also displayed for comparison.
We can see that compared with the DM self-interaction model \cite{Tulin:2013teo},
 the GMSSM has a larger parameter space to solve the anomalies of
 all three small scales (the mass of DM can be heavier than 10 GeV and
the coupling strength $\alpha_\chi$ can be at order $10^{-3}$).
The reason is that in the DM self-interaction model \cite{Tulin:2013teo}
DM can only annihilate into $hh$ via $t$-channel and $u$-channel while in the GMSSM
DM can annihilate into $hh$, $ha$ and $aa$ via $t$-channel, $u$-channel and $s$-channel,
as shown in Fig.\ref{fig9}.

\begin{figure}[htbp]
\scalebox{0.8}{\epsfig{file=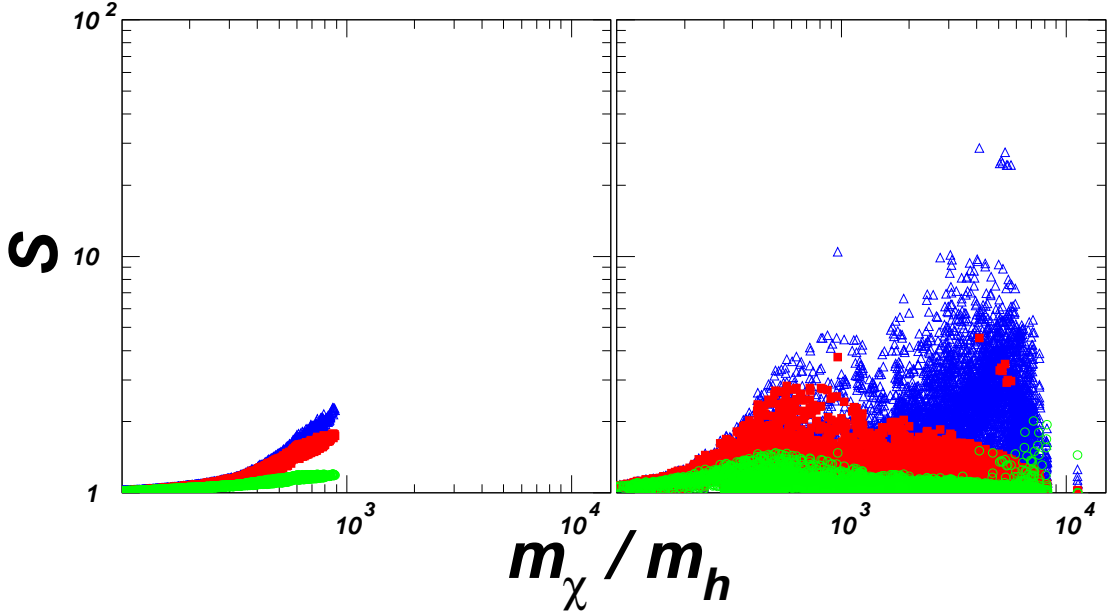}}
\caption{Same as Fig.\ref{fig10}, but showing
the Sommerfeld enhancement factor.
The left frame is for the DM self-interaction
model \cite{Tulin:2013teo}
while the right frame is for the GMSSM.}
\label{fig12}
\end{figure}
Finally, in Fig. \ref{fig12} we show the Sommerfeld enhancement in the
allowed parameter space.
We see that compared with the DM self-interaction model \cite{Tulin:2013teo},
the GMSSM allows for a larger $m_\chi/m_h$  and thus can give a larger
Sommerfeld enhancement factor.
The ongoing indirect detections of DM can probe the
DM annihilation rate (shed light on the Sommerfeld enhancement factor) and
thus help to distinguish different DM self-interaction models.

\section{Conclusions}\label{sec5}
In this paper we studed the possibility of  the DM self-interaction to
solve the small cosmological scale anomalies
in the singlet extensions of the MSSM.
We first checked the NMSSM and found that
the correlation between the DM annihilation rate and DM-nucleon SI cross section
strongly constrains this model so that it cannot  realize the DM self-interacting
scenario.
For the GMSSM, the parameter space was found to be large enough to
realize the DM self-interacting scenario and at the same time can give
a large Sommerfeld enhancement factor.
Also, we found that for Majorana-fermion DM, we must use viscosity cross sections
($\sigma_{VS}$ and $\sigma_{VA}$)
in DM simulations.

\section*{Acknowledgments}
This work was supported by the
Natural Science Foundation of China under grant numbers 11105124,11105125,
11275245, 10821504,  11135003, 11005006, 11172008, 11375001
 and Ri-Xin Foundation of BJUT.

\end{document}